**Title** High-power-handling ultra-compact acousto-optic modulators using one-dimensional topological interface states on thin-film lithium tantalate

*Yuqi Chen,[1,2] Wenfeng Zhou,[1,2] Min Sun,[1] Xun Zhang,[1] Xin Wang,[1] Qingqing Han,[1] Minni Qu,[1] Yikai Su,[1] and Yong Zhang[1,*]*
*Corresponding Author: E-mail: yongzhang@sjtu.edu.cn

[1]State Key Laboratory of Photonics and Communications, School of Integrated Circuits, Shanghai Jiao Tong University, Shanghai 200240, China
[2]These authors contributed equally.

Recent advances in integrated photonics have enabled on-chip signal modulation and processing through localized photon-phonon interactions. For acousto-optic devices, compact footprint and high efficiency are essential for dense integration, while strong power handling is critical for stable operation in demanding applications. However, it remains challenging to achieve these features simultaneously on existing integrated platforms. Here, we propose and experimentally demonstrate, for the first time on a thin-film lithium tantalate platform, an ultra-compact acousto-optic modulator based on topological interface states. Benefiting from the strong optical confinement of the topological boundary state, the device achieves a footprint of $130 \times 120$ μm$^2$ and a half-wave voltage-length product of 0.491 V·cm. We further demonstrate stable acousto-optic modulation at an on-chip optical power of up to 28 dBm (630.9 mW), highlighting the strong power-handling capability of the thin-film lithium tantalate topological structure. This work provides a compact and high-power solution for microwave-to-photonic transduction and shows the potential of the thin-film lithium tantalate for robust integrated photonic systems.

# 1. Introduction

Acousto-optic (AO) interaction represents a fundamental multiphysics coupling process, wherein radio-frequency (RF)-driven acoustic waves modulate the local refractive index of a waveguide to enable precise control over photonic states[1-4]. AO devices have found extensive applications in optical modulation[5-6], frequency shifting[7-8], microwave signal processing[9-10], optical computing[11-12], and quantum information science[13-14]. Traditional bulk AO modulators, typically based on materials such as quartz and lithium niobate (LN), exhibit high modulation efficiencies; however, they are inherently limited by their bulky footprints, restricted scalability, and fundamental incompatibility with photonic integrated circuits (PICs)[15]. In contrast to bulk platforms, PICs facilitate strong confinement of surface acoustic waves (SAWs) within thin films, significantly enhancing the overlap between the acoustic and optical fields and making SAWs the basis of most high-performance on-chip AO architectures[16]. By exploiting piezoelectric actuation and the photoelastic effect, integrated AO modulators have been successfully demonstrated across a variety of material platforms, including lithium niobate[17], aluminum nitride (AlN)[18], aluminum scandium nitride (AlScN)[19], gallium arsenide (GaAs)[20], and zinc oxide (ZnO)[21].

Owing to its excellent electro-optic and piezoelectric properties, thin-film lithium niobate (TFLN) has become a leading platform for high-performance PICs, and its rapid development has also driven major progress in integrated AO devices, with important applications in data center interconnects[22-23], high-performance computing[24-25], and artificial intelligence[26]. Currently, high-performance AO modulators are primarily realized through Mach-Zehnder interferometers (MZIs) or microring resonators[27-28]. By integrating suspended acoustic cavities with TFLN racetrack resonators, researchers have achieved a state-of-the-art half-wave voltage–length product $V_\pi \cdot L$ as low as 0.0077 V·cm—a nearly two-order-of-magnitude improvement in efficiency compared to non-suspended counterparts[29]. Furthermore, heterogeneous integration of chalcogenide glass (ChG) with TFLN in push-pull configurations has pushed $V_\pi \cdot L$ down to 0.03 V·cm[30]. However, this pursuit of extreme modulation efficiency often compromises high-power operational robustness. High-efficiency devices typically rely on suspended architectures that suffer from mechanical instability and limited thermal dissipation, while undoped lithium niobate itself exhibits pronounced photorefractive effects under high optical power, leading to bias-point drift and reduced stability[31-33]. These limitations are particularly critical in applications such as microwave generation, nonlinear frequency conversion, and high-linearity analog photonic links, all of which demand superior power-handling capabilities[34-38]. In addition, although TFLN is a promising platform for integrated photonics, its broader use in large-scale applications may still require further progress in fabrication cost and manufacturing maturity. Consequently, developing a novel material platform that harmonizes exceptional acousto-optic performance with superior power tolerance has become a critical imperative for the advancement of robust integrated photonics.

The thin-film lithium tantalate (TFLT) platform presents a compelling alternative, leveraging its unique advantages in microwave-photonic integration. Unlike TFLN, TFLT has already achieved extensive industrial adoption in 5G RF front-end components (e.g., surface acoustic wave filters), underscoring its superior material maturity and scalable manufacturability[39]. Regarding its optical properties, the birefringence of TFLT ($\Delta n = 0.004$) is more than an order of magnitude lower than that of TFLN ($\Delta n = 0.07$)[40], effectively suppressing parasitic mode coupling and stray scattering induced by refractive index mismatch, thereby enhancing design flexibility and link fidelity in high-density photonic integration. Moreover, TFLT exhibits an optical damage threshold significantly higher than that of TFLN while maintaining low propagation losses[41], making TFLT an ideal platform for stable, high-performance acousto-optic devices operating in power-intensive regimes. Despite these advantages, efficient on-chip acousto-optic modulators on the TFLT platform remain largely unexplored[42].

Here, we propose and experimentally demonstrate the first integrated AO modulator on a TFLT platform, leveraging 1D topological interface states. Inspired by the classic 1D Su–Schrieffer–Heeger model, a topologically protected boundary state (TBS) is formed at the interface between two photonic crystal nanobeam lattices with distinct topological invariants. Using SAW generated by an interdigital transducer (IDT), the device enables efficient acousto-optic interaction within a compact footprint of 130 × 120 μm$^2$ and achieves a half-wave voltage–length product V$\pi$·L of 0.491 V·cm. We further demonstrate stable operation at optical powers up to 28 dBm (630.9 mW), indicating a high optical damage threshold and suppressed PR effects. These results highlight the potential of combining the material advantages of TFLT with topological states for high-power, multifunctional integrated photonic systems.

## 2. Device Design and Operation Principle

**Figure** 1a illustrates the vision of a monolithically integrated multifunctional PIC built on the TFLT platform. In this architecture, the on-chip AO modulator serves as a front-end electro-acousto-optic interface, where high-power optical input and microwave signals are coupled to realize coherent microwave-to-photonic transduction before routing the signal to backend photonic circuits. The modulated optical signal can then be directed to different functional modules for microwave photonic signal processing, lidar-based beam steering, and high-speed optical convolution computing. Its inherent frequency up-conversion capability enables compact and efficient transfer of microwave signals into the optical domain, making it well suited for integration with subsequent multifunctional photonic processing units. By taking advantage of the scalability and high optical power tolerance of TFLT, such an architecture provides a promising route toward high-capacity, power-robust, and cost-effective integrated photonic systems.

To implement the front-end AO modulation function in the integrated architecture of Figure 1a, we propose and demonstrate a high-efficiency AO modulator, as shown in Figure 1b. Its core is a one-dimensional topological photonic crystal (1D-TPC) nanobeam cavity defined by a periodic array of air holes on an X-cut TFLT wafer. The wafer consists of a 600-nm-thick LT thin film and a 9-μm-thick buried oxide layer on a silicon substrate. The LT ridge waveguide features an etch depth of 300 nm, a width of 1.2 μm, and a sidewall angle of approximately 85°. Spatially, the optical waveguide is aligned along the crystal Y-axis, while the aperture of the IDT is aligned with the Z-axis. Upon driving the IDT with a microwave signal, Rayleigh surface acoustic waves propagating along the Z-axis are excited, enabling efficient modulation of the topologically localized optical field through the generalized acousto-optic effect. More detailed information about the generalized acousto-optic effect can be found in Section S1 (Supporting Information).

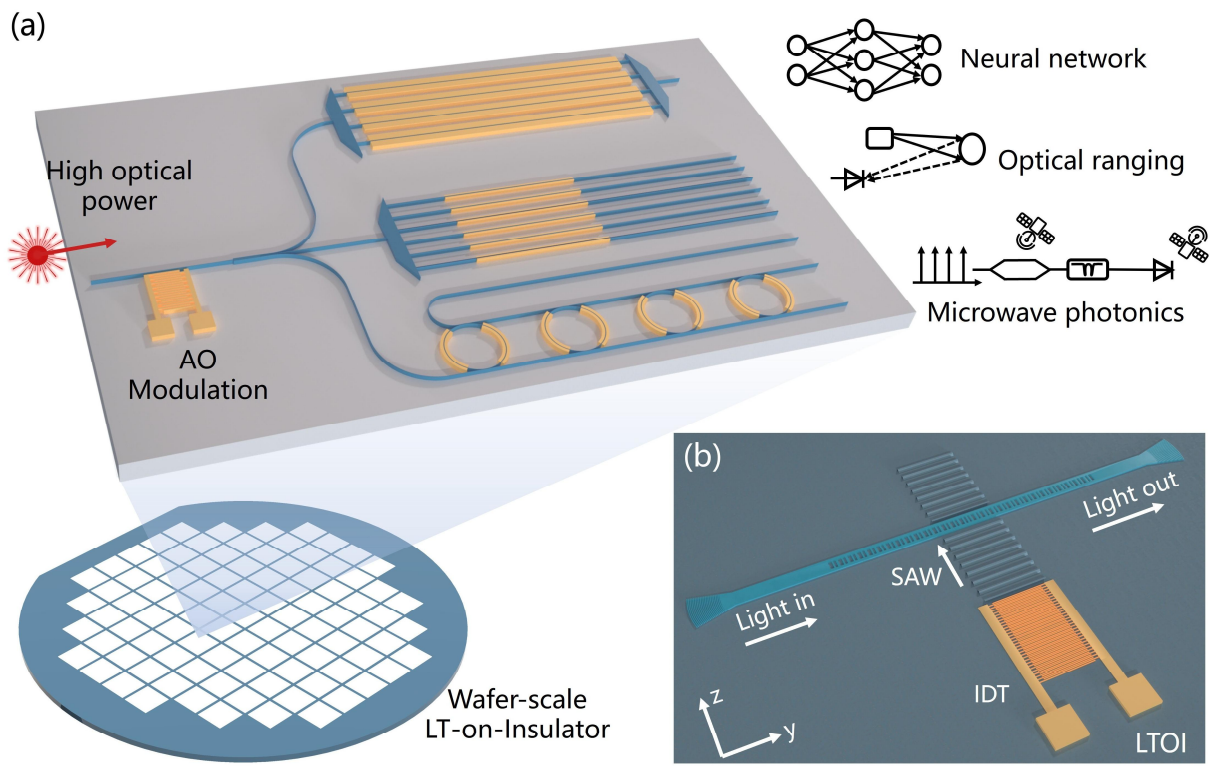

**Figure 1**. Integrated TFLT photonic circuits: from large-scale system concepts to individual functional devices. (a) Schematic of a next-generation large-scale PIC based on the TFLT platform, featuring high optical power handling and comprising an integrated acousto-optic modulation block with a low-loss, multipurpose photonic processing unit. (b) Schematic of the one-dimensional topological interface state–based integrated AO modulator on the TFLT platform.

Based on the classic SSH model, we design a 1D-TPC nanobeam cavity by joining two nanobeam lattices with distinct topological properties. To preserve spatial inversion symmetry, the two lattices are arranged in a mirror-symmetric manner about the cavity center, as shown in **Figure** 2a. Each unit cell has a lattice constant Λ of 411 nm and contains two rectangular air holes with the same length (D3 = 312 nm) and different widths (D1 = 176 nm and D2 = 116 nm), which modulate the local effective refractive index and emulate the alternating dielectric distribution in the SSH model. Although the two nanobeam lattices exhibit identical photonic band structures, they possess fundamentally different topological invariants. These are characterized by the Zak phase, which is defined as:

$$\theta_{Zak} = \int_{BZ}\left[ i\int d\gamma \cdot \varepsilon(\gamma) u_k^*(\gamma) \partial_k u_k(\gamma) \right] dk , \qquad (1)$$

where BZ denotes the first Brillouin zone, $\varepsilon(\gamma)$ represents the dielectric constant distribution, and $u_k(\gamma)$ is the normalized Bloch function.

Under spatial inversion symmetry, the Zak phase is quantized to either 0 or π. As shown in Figure 2c,d, exchanging the positions of the two air holes does not change the band structure, but reverses the modal symmetry at the band edges, indicating a band inversion between the two lattices. As a result, although the two lattices share the same photonic bandgap, they possess different Zak phases. When these two topologically distinct lattices are joined together, a localized interface state is formed inside the common bandgap. As shown in the upper panel of Figure 2e, this state gives rise to a sharp resonance peak within the bandgap, corresponding to the topological boundary state (TBS). After parameter optimization, the topological cavity exhibits an extinction ratio of 15 dB and an insertion loss of about 0.76 dB. Moreover, the TBS is spectrally separated from the band-edge modes, which helps reduce mode competition and improves robustness against structural perturbations [43]. Further technical details employed in the device design are provided in Section S2 (Supporting Information).

During device operation, the IDT is driven by a microwave signal to generate SAWs that travel along the TFLT surface and interact with the topological cavity. This interaction changes the effective refractive index of the cavity mode and causes optical power modulation in transmission. At a given bias point, the modulation amplitude is proportional to the local slope of the optical transmission spectrum. Based on this relation, the corresponding photoacoustic $S_{21}$ response can be extracted from the simulated transmission modulation, providing a quantitative measure of the photon-phonon interaction, as shown in the lower panel of Figure 2e. The results indicate that the $S_{21}$ response attains its maximum at a laser detuning point approximately 3 dB below the peak non-resonant transmission level. This observation is consistent with the slope-detection mechanism, as this specific detuning corresponds to the region where the transmission curve exhibits its steepest gradient ($dT/d\lambda$). At this optimal bias point, the spectral shifts induced by the SAWs are most efficiently transduced into optical intensity fluctuations, thereby maximizing the acousto-optic modulation efficiency.

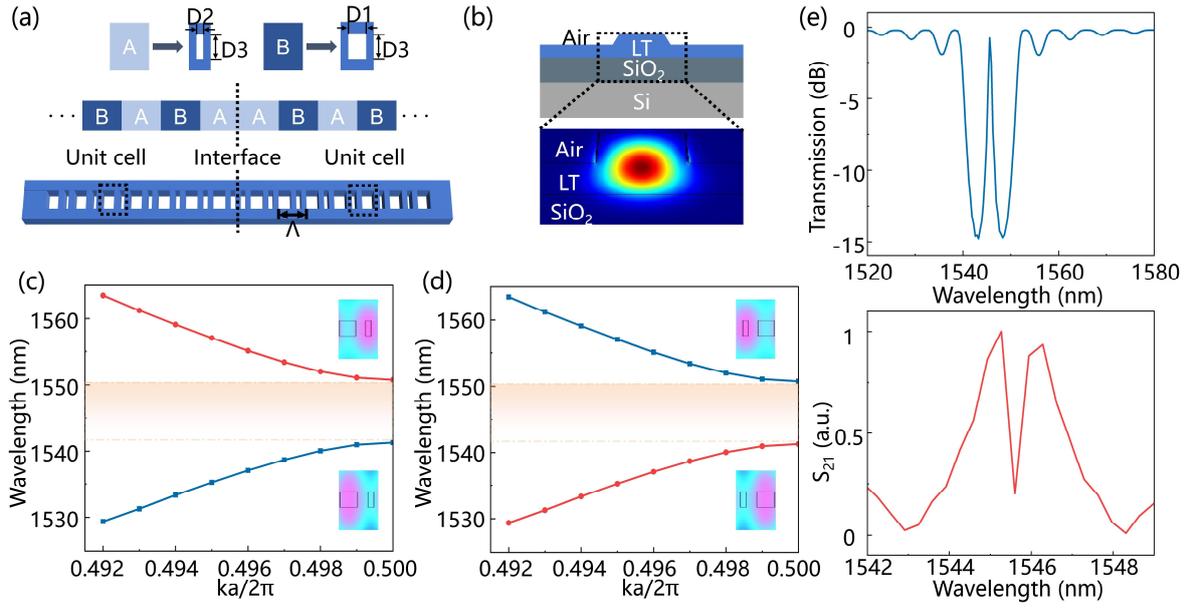

**Figure 2**. Structural design and optical characterization of the TFLT topological photonic crystal cavity. (a) Detailed schematic diagram of a 1D-TPC nanobeam cavity. (b) Cross-sectional schematic of the etched LT ridge waveguide and optical field distribution of the fundamental TE-like mode, with an effective refractive index of 1.91 at 1550 nm. Band structures of the (c) left and (d) right integrated TPC cavity based on periodic rectangular holes. (e) Simulated optical transmission and acoustic-perturbed $S_{21}$ response evolution of the integrated TPC cavity.

To evaluate the device performance and optimize the structural parameters, eigenmode analysis is used to study the acoustic modes supported by the IDT. **Figure** 3a illustrates the top view of the IDT electrode configuration, where L, a, and N denote the AO modulation aperture, the finger width, and the number of finger pairs, respectively. The optimized IDT comprises 30 finger pairs with a pitch of 4.4 μm, corresponding to an electrode width of 1.1 μm. In the eigenmode simulations, a reduced number of electrode pairs is employed as a representative subset to accurately reflect the acoustic mode distribution while maintaining computational efficiency. Figure 3b displays the numerical simulation of the primary $S_{xx}$ strain component for the SAW mode at 0.54 GHz. The acoustic waves propagate along the TFLT surface and interact with the ridge waveguide, inducing mechanical deformation. Furthermore, the proposed IDT is capable of exciting higher-order acoustic modes. Figure 3c presents the eigenmode profiles of different acoustic frequencies for a single electrode pair, clearly showing that adjacent electrodes exhibit antiphase deformation at 0.54 GHz. Notably, the acousto-optic overlap factor for higher-frequency modes is significantly diminished, as a larger fraction of the acoustic energy leaks into the substrate rather than contributing to the photon-phonon interaction.

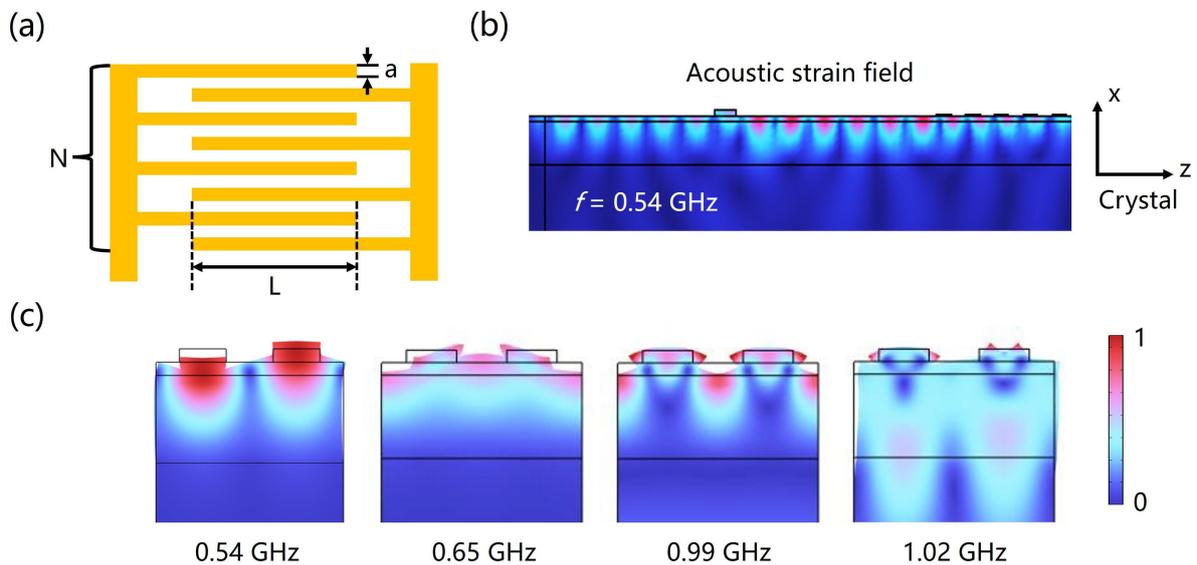

**Figure 3**. Design and acoustic mode analysis of the IDT on the TFLT platform. (a) Top-view schematic of the IDT electrode configuration with defined geometric parameters (L, a, and N). (b) Distribution of the acoustic strain field ($S_{xx}$) induced by the 0.54 GHz SAW mode. The strain magnitudes are normalized to the peak frequency response. (c) Normalized displacement distributions of various acoustic modes, highlighting the antiphase deformation at 0.54 GHz and the significant substrate leakage associated with higher-order modes.

## 3. Device Fabrication and Characterization

The TPC-AO modulator is fabricated on a commercial X-cut TFLT wafer (Inno Semiconductor Co., Ltd.). To establish a robust etching framework, a 600-nm-thick amorphous silicon (a-Si) hard mask is first deposited via plasma-enhanced chemical vapor deposition (PECVD), followed by the application of AR-P 6200.09 photoresist through spin-coating. The device structures—including the 1D topological lattices, optical waveguides, and grating couplers—are defined on the resist using electron-beam lithography (EBL, Vistec EBPG-5200+). Subsequently, the patterns are transferred into the a-Si hard mask through inductively coupled plasma (ICP) etching. Utilizing the patterned a-Si layer, the underlying LT film is etched to a depth of 300 nm. The residual hard mask is then stripped by immersing the sample in a KOH:water solution, yielding the finalized on-chip LT photonic structures. The fabrication concludes with the definition of 300-nm-thick gold (Au) electrodes through electron-beam evaporation and a standard lift-off process, resulting in the on-chip lithium tantalate AO modulators. More details of the device fabrication process can be found in Section S3 (Supporting Information).

The fabricated device is characterized using both optical microscopy and scanning electron microscopy (SEM), as presented in **Figure** 4a–c. To facilitate high-frequency signal delivery via ground-signal-ground (G-S-G) RF probes, the IDTs are integrated with a three-pad contact configuration. Specifically, two pads are utilized to drive the IDT for SAW generation, while the third pad ensures a stable connection for the probe's ground pin, preventing it from remaining floating. Figure 4b provides a magnified view of the 1D topological interface state, showcasing the high-fidelity etching of the air-hole lattices within the waveguide. To evaluate the optical and acousto-optic performance, on-chip grating couplers are implemented to enable efficient fiber-to-chip coupling of the fundamental TE mode. The measured coupling loss for each grating coupler is approximately 7 dB at the target operating wavelength (Section S4, Supporting Information).

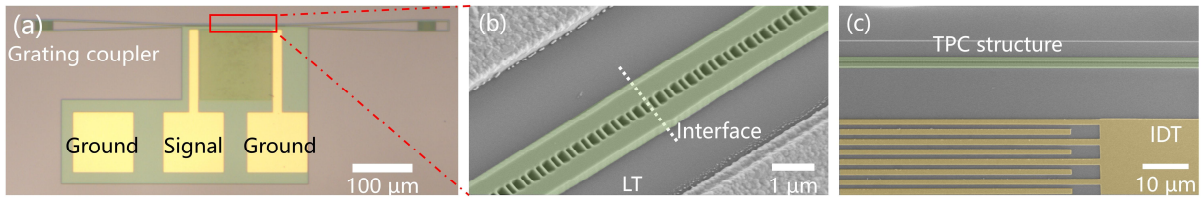

**Figure 4**. Configuration of the integrated acousto-optic modulator based on a TPC cavity over the non-suspended TFLT platform. (a) Optical microscopy image of the fabricated device. (b) Detailed SEM view of the 1D topological interface, illustrating the precise etching of the photonic crystal nanobeam. (c) Top-view SEM image of the TPC-AO modulator, with the IDT electrode structure located on one side of the waveguide. False color is applied to the waveguide and IDT electrodes for clarity.

**Figure** 5a illustrates the schematic of the experimental configuration for characterizing the on-chip AO modulator. A C-band tunable laser source (Santec TSL-570) provides the input light, which is subsequently adjusted by a polarization controller (PC) to ensure operation in the fundamental TE mode. At the output of the TPC-AO modulator, a 3-dB coupler splits the signal: 50% is directed to an optical spectrum analyzer (OSA) for characterizing the static optical resonance of the topological cavity, while the remaining 50% is routed to an erbium-doped fiber amplifier (EDFA) to compensate for transmission losses. The amplified optical signal is further regulated by a variable optical attenuator (VOA) before being captured by a photodetector (PD) with a sensitivity of 29.5 V/W. For dynamic characterization, the drive port of a Vector Network Analyzer (VNA) (ZNA67) is connected to the IDT to excite surface acoustic waves, while its receiver port records the modulated signal from the PD. The dashed box in Figure 5a delineates the modified test link designed for high-power input conditions. In this configuration, high

optical power levels are generated by combining the tunable laser with an EDFA. Due to the sufficiently high output power in this regime, the subsequent EDFA-based loss compensation is bypassed.

The experimental transmission spectrum of the topological cavity is presented in Figure 5b, revealing a sharp resonance peak located at 1543.5 nm, positioned precisely at the center of the photonic bandgap. The fabricated TPC device exhibits an extinction ratio of 17 dB, an insertion loss of 1.53 dB, and a quality factor Q of 4244.

To evaluate the microwave-to-acoustic transduction efficiency, we measure the microwave reflection coefficient ($S_{11}$) spectrum of the fabricated IDT. As illustrated in Figure 5c, the spectrum exhibits four distinct resonance minima within the 0.1–1.15 GHz range, corresponding to four acoustic modes at 0.54 GHz, 0.65 GHz, 0.99 GHz, and 1.02 GHz (labeled as Modes 1, 2, 3, and 4, respectively). These experimental results are in excellent agreement with our numerical simulations. The minor frequency shifts observed are likely attributable to fabrication tolerances and the simplified electrode count used in the simulation model. For Mode 1, a deep resonance dip occurs at 0.544 GHz with a 1-$|S_{11}|^2$ value of 83.1%, which represents the ratio of RF power effectively loaded into the IDT relative to the input power. The acoustic quality factor (Q_acoustic) of this resonance is estimated to be approximately 413.

The AO modulation performance is further characterized via the opto-acoustic $S_{21}$ transmission spectrum. To maximize the modulation sensitivity, the laser wavelength is tuned to a 3-dB detuning point on the resonance slope, as established in the preceding section. The on-chip optical input power is consistently set to 0 dBm throughout the measurement. The $S_{21}$ spectrum in Figure 5d displays two prominent peaks between 0.1 GHz and 1.15 GHz, indicating enhanced microwave-to-optical conversion at these specific frequencies. At an RF drive power of 0 dBm, a robust AO response is observed at 0.544 GHz, correlating perfectly with the sharp $S_{11}$ dip of Mode 1. Interestingly, although Mode 3 exhibits a higher 1-$|S_{11}|^2$ value of 91.5%—suggesting superior microwave power absorption—its corresponding $S_{21}$ magnitude (–69.01 dB) is significantly weaker than that of Mode 1 (–59.91 dB). This discrepancy confirms that Mode 1 possesses a substantially larger acousto-optic overlap integral with the TPC cavity mode, leading to superior modulation efficiency despite the slightly lower electrical impedance matching compared to Mode 3.

To further evaluate the acousto-optic modulation efficiency, the half-wave voltage ($V_\pi$) of the proposed non-suspended modulator is derived from the experimental $S_{21}$ response utilizing the following relationship[27]:

$$V_\pi = \frac{\pi R_{PD} I_{\text{rec}}}{|S_{21}|}, \qquad (2)$$

where $R_{PD}$ is the sensitivity of the photoreceiver and $I_{rec}$ is the DC optical power at the bias point. By selecting the bias point on the transmission spectrum, a received optical power $I_{rec}$ of -3.5 dBm is recorded. Integrating this with the $S_{21}$ response at 0.544 GHz, the half-wave voltage-length product ($V_\pi \cdot L$) is determined to be 0.491 V·cm for the 120-μm-long modulation region. The signal-to-noise ratio (SNR) for the primary modulation peak at 0.544 GHz exceeds 25 dB, highlighting the clean signal transduction of the TPC-based architecture. As the first AO modulation reported on TFLT, the device exhibits a high efficiency of 0.49 V·cm, a performance level that approaches the top-tier metrics of TFLN-based counterparts and fills a critical gap in TFLT integrated acousto-optics.

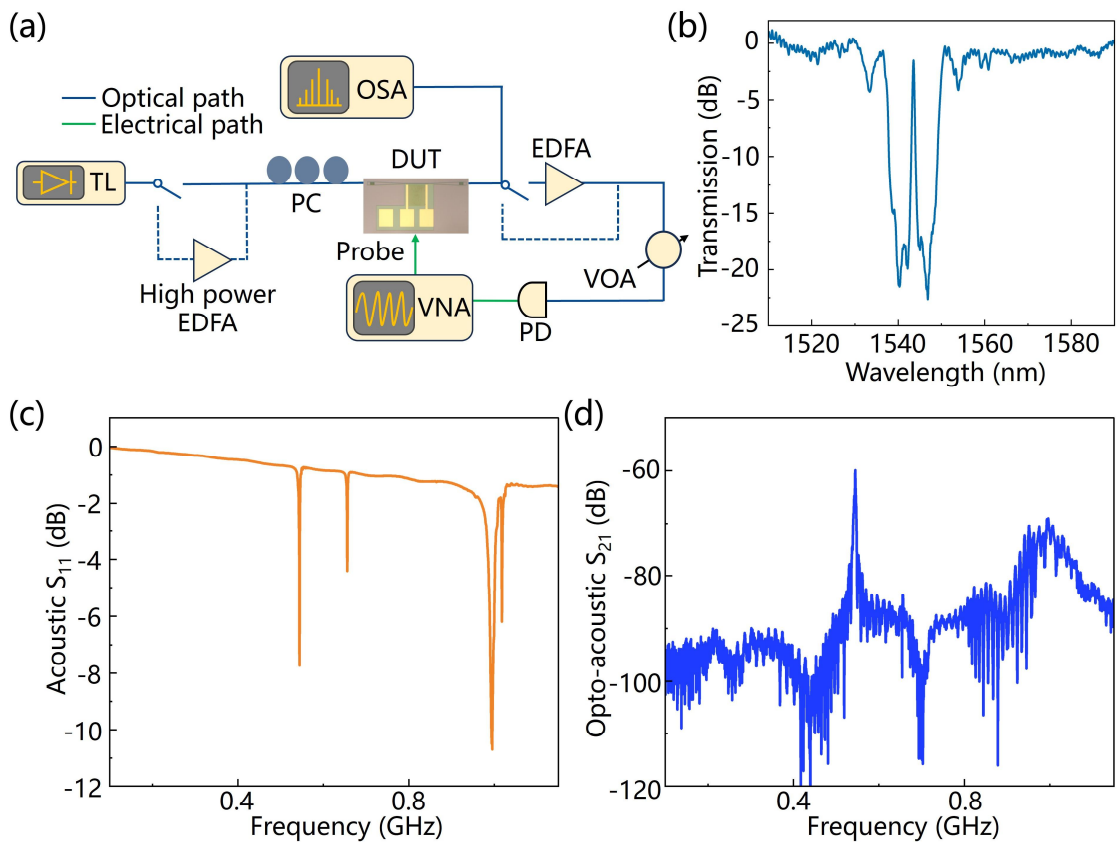

**Figure 5**. Experimental characterization of the on-chip TPC acousto-optic modulator. (a) Schematic diagram of the measurement experimental setup (TL: tunable laser; PC: polarization controller; EDFA: erbium-doped fiber amplifier; DUT: device under test; VOA: variable optical attenuator; PD: photodetector; VNA: vector network analyzer; OSA: optical spectrum analyzer). (b) Transmission spectrum of the device; the inset shows the Lorentzian fit. (c) $S_{11}$ reflection spectrum of the fabricated IDT. (d) Measured opto-acoustic $S_{21}$ spectrum of the TPC-AO modulator.

## 4. High-Power Stability of the Acousto-Optic Modulation

Beyond modulation efficiency, the power-handling capability of an on-chip AO modulator is a key metric for applications, particularly in integrated photonic systems that require stable acousto-optic modulation under high optical power, such as high-power transmitting modules for lidar, remote sensing, space optical communication, and laser ranging[44-46]. To evaluate this capability, we measure the stability of the proposed TPC-AO modulator at input powers of 25, 30, and 35 dBm. With a coupling loss of about 7 dB per grating coupler, the corresponding on-chip powers are 18, 23, and 28 dBm, respectively.

First, we characterize the static optical response of the TPC cavity at different on-chip optical power levels. As shown in **Figure** 6a–c, the resonance wavelength in the transmission spectra remains nearly unchanged as the optical power in the waveguide increases from 18 dBm to 28 dBm, indicating good thermo-optic stability of the non-suspended TFLT platform. However, a noticeable broadening of the resonance linewidth is observed at higher power levels. This phenomenon is likely attributable to the intrinsic nonlinearities of the LT crystal, such as multi-photon absorption or nonlinear refractive index effects, which become prominent under intense localized field enhancement within the topological cavity. Despite the high on-chip optical power, the insertion loss remains nearly unchanged, while the extinction ratio shows only a slight decrease, in good agreement with the simulations. These results confirm that the TPC cavity maintains robust optical confinement and structural integrity even in the high-power regime, providing a solid basis for stable acousto-optic modulation.

Furthermore, we evaluate the acousto-optic modulation performance at different on-chip optical power levels. Figure 6d–f shows the measured $S_{11}$ and $S_{21}$ spectra at on-chip powers of 18, 23, and 28 dBm. As expected, the $S_{11}$ spectra remain nearly identical in all cases, with the four acoustic modes consistently showing resonance dips at their characteristic frequencies, indicating that the microwave reflection response of the IDT is independent of optical power. To keep the PD in the linear regime, a VOA is used to control the optical power reaching the PD. The measured $S_{21}$ values at 0.544 GHz are −49.91, −49.85, and −50.17 dB, respectively, corresponding to received optical powers $I_{rec}$ of 1.5, 1.5, and 1.4 dBm. According to Equation 2, the extracted $V_\pi \cdot L$ values are 0.492 V·cm, 0.488 V·cm, and 0.495 V·cm, respectively. The maximum variation relative to the low-power case is less than 0.004 V·cm, indicating that the microwave-to-optical transduction remains essentially unchanged and confirming the mechanical and optical stability of the device under high optical power.

To assess the long-term stability, we perform a continuous test at an on-chip optical power of 28 dBm. Figure 6g shows the temporal evolution of the resonance wavelength and relative output intensity. During the measurement, the resonance wavelength remains stable, while the output intensity fluctuation stays within 0.6 dB. As shown in Figure 6h, the $S_{21}$ magnitude at 0.544 GHz and the corresponding Vπ·L also remain stable over 30 minutes. In addition, the SNR of the main modulation peak remains above 25 dB throughout the measurement. These results confirm the robustness of the device under high-power operation. Minor fluctuations observed at several time points are likely caused by small mechanical disturbances in the fiber-to-chip coupling under high thermal load. Detailed stability characterization data are provided in Section S5 (Supporting Information). It is noteworthy that the maximum input power characterized in this study is strictly limited by the saturation output of the available laser source and EDFA, rather than the intrinsic damage threshold of the TFLT device.

Our findings indicate that TFLT waveguides exhibit a markedly suppressed PR effect and a higher optical damage threshold compared to their TFLN counterparts. This superior power-handling capability is likely rooted in the intrinsic material properties of lithium tantalate, including its higher resistance to photon-induced displacement, the superior strength of the Ta–O chemical bonds, and a significantly lower density of intrinsic vacancies and defect states[33]. The stable operation at an internal optical power of 28 dBm, representing the highest reported value so far for integrated AO modulators, highlights the strong power-handling capability of our TPC-based architecture.

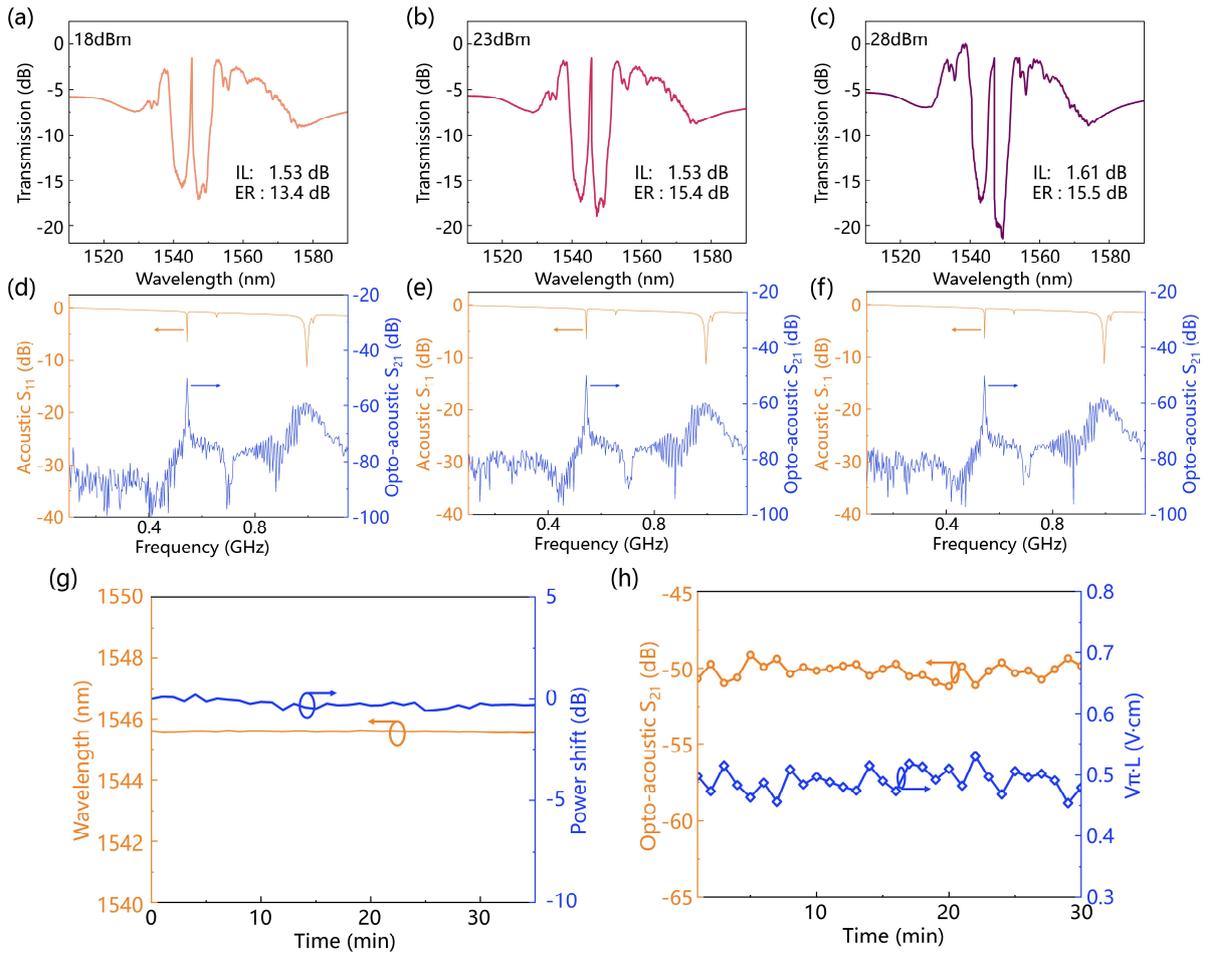

**Figure 5**. High-power stability of the TPC-AO modulator. Transmission spectra of the TPC-AO modulator at on-chip optical powers of (a) 18 dBm, (b) 23 dBm, and (c) 28 dBm. Measured acoustic $S_{11}$ and opto-acoustic $S_{21}$ spectra at on-chip optical powers of (d) 18 dBm, (e) 23 dBm, and (f) 28 dBm. (g) Temporal variations of the resonance wavelength and relative output intensity at an on-chip optical power of 28 dBm. (h) Stability test of the AO modulation efficiency at an on-chip optical power of 28 dBm.

**Table 1** presents a performance comparison between our 1D-TPC AO modulator on the TFLT platform and representative integrated AO devices based on various TFLN structures. While modulation efficiency is a primary metric, device footprint also plays an essential role in achieving high-density integration. Traditional MZI- or MRR-based AO modulators necessitate additional building blocks such as beam splitters, directional couplers, or large-radius bends, which significantly increase their footprint. In contrast, by harnessing the strong optical confinement of the topological interface state, our TPC-AO modulator accomplishes efficient modulation within a compact area of only 130 × 120 μm²—roughly one order of magnitude smaller than that of conventional integrated AO modulators. Notably, this study demonstrates the first integrated AO modulator on the TFLT platform and, to the best of our knowledge, the first experimental validation of stable AO modulation under high on-chip optical power. The device sustains stable operation at an internal optical power as high as 28 dBm (630.9 mW), which corresponds to an approximately two-order-of-magnitude enhancement in power handling capability relative to prior integrated AO devices.

**Table 1**. Performance of various state-of-the-art integrated AO modulators.

| Platform | Modulator structure | Acoustic cavity | Frequency (GHz) | Modulation length (μm) | $V_\pi \cdot L$ (V·cm) | Footprint (mm²) | Optical Power (mW) |
|---|---|---|---|---|---|---|---|
| LN | MZI[28] | √ | 0.11 | 1200 | 2.5 | 1.2 × 5.7[a] | / |
| LN-As$_2$S$_3$ | MZI[47] | √ | 0.11 | 2400 | 0.94 | 1.2 × 5.2[a] | / |
| LN-ChG | MZI[30] | × | 0.84 | 120 | 0.03 | 2 × 3[a] | 0.224[c] |
| LN-Si | MZI[48] | × | 0.47 | 1000 | 0.50 | 0.4 × 1[b] | / |
| LN | Racetrack[29] | √ | 2.17 | 100 | 0.0077 | 0.4 × 0.4[b] | 0.820[c] |
| LN-ChG | Racetrack[49] | × | 0.84 | 120 | 0.02 | 2 × 2[a] | / |
| LN-ChG | π-PSBG[50] | × | 0.84 | 300 | 0.03 | 0.2 × 0.3[a] | 0.631[c] |
| LT (this work) | TPC | × | 0.54 | 120 | 0.49 | 0.12 × 0.13 | 630.9 |

[a] Results are from the literature;

[b] Estimated from the SEM figure;

[c] Estimated from the data provided in the literature.

**5. Conclusion**

We demonstrate the first on-chip acousto-optic modulator on the TFLT platform based on a one-dimensional TPC cavity. By using the strong optical confinement of the topological interface state, the device achieves a half-wave voltage-length product Vπ·L of 0.491 V·cm within a compact footprint of 130 × 120 μm$^2$. Compared with conventional MZI- and microring-based AO modulators, this design offers a much smaller footprint and is therefore well suited for dense photonic integration. We also show stable AO modulation at an on-chip optical power of up to 28 dBm (630.9 mW), which is, to our knowledge, the first such demonstration for an integrated AO device. This result highlights the strong power-handling capability of the TFLT platform and reflects the benefits of both the material properties of LT and the non-suspended device structure. Overall, this work shows that topological cavity enhancement on TFLT is a promising approach for realizing compact, efficient, and robust AO devices. Further improvements in the piezoelectric transducer and cavity design may reduce the Vπ·L to below 0.1 V·cm and support future high-power acousto-optic signal processing and nonlinear photonic circuits.

**Supporting Information**

Additional supporting information may be found in the online version of this article at the publisher's website.

**Acknowledgements**

This work was supported in part by the National Natural Science Foundation of China (NSFC) under Grant 62335014 and the Shanghai Municipal Science and Technology Commission Project under Grant 25JD1402000. We thank the Center for Advanced Electronic Materials and Devices (AEMD) of Shanghai Jiao Tong University (SJTU) for their support in device fabrication.

**Graphical Abstract**
In this study, we report the first demonstration of a high-performance acousto-optic modulator on the thin-film lithium tantalate (TFLT) platform, surpassing the power-handling constraints of conventional lithium niobate. By pioneering the integration of 1D topological interface states, we achieve robust optical confinement within an ultra-compact footprint (130 × 120 μm²). The device yields a remarkable Vπ·L of 0.491 V·cm and maintains stable modulation under the high optical power up to 28 dBm. This work establishes TFLT-based topological architectures as a robust, scalable solution for next-generation microwave-to-photonic conversion in demanding, high-power integrated environments.

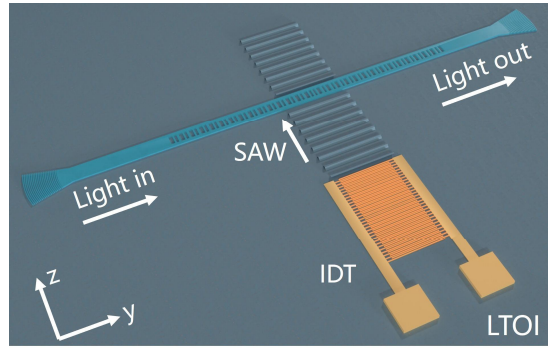